\documentstyle[12pt]{article}

\newcommand{\sPP}{\mathrm{I}\kern -1.6pt \mathrm{P}}
\newcommand{\sR}{\mathrm{I}\kern -1.7pt \mathrm{R}}
\newcommand{\sZ}{\mathrm{Z}\!\!\mathrm{Z}}

\def\th{\theta}

\def\a{\alpha}
\def\b{\beta}
\def\d{\delta}

\def\l{\lambda} \def\L{\Lambda}

\def\s{\sigma}

\def\longrightarrow{\relbar\joinrel\relbar\joinrel\rightarrow}
\def\be{\begin{equation}}
\def\ee{\end{equation}}

\renewcommand{\baselinestretch}{1.0}

\begin{document}

\begin{flushright}
BRX TH-534
\end{flushright}

\vspace*{1in}

\begin{center}
{\Large\bf Confinement/Deconfinement Transition of \\Large
$\mbox{\boldmath$N$}$ Gauge Theories with $\mbox{\boldmath$N_f$}$
Fundamentals:\\ $\mbox{\boldmath$N_f/N$}$ Finite}

\vspace{.15in}

\renewcommand{\baselinestretch}{1}
\small
\normalsize

Howard J. Schnitzer\footnote{email: schnitzr@brandeis.edu}\\
Martin Fisher School of Physics\\
Brandeis University\\
Waltham, MA 02454

\vspace{.2in}

 {\bf Abstract}
\end{center}

\begin{quotation}
We consider large $N$ zero-coupling $d$-dimensional U(N) gauge
theories, with $N_f$ matter fields in the fundamental
representation on a compact spatial manifold $S^{d-1}\; \times$
time, with $N_f/N$ finite.  The Gauss' law constraint induces
interactions among the fields, in spite of the zero-coupling. This
class of theories undergo a 3$^{\rm rd}$ order deconfinement phase
transition at a temperature $T_c$ proportional to the inverse
length scale of the compact manifold.\\

The low-temperature phase has a free-energy of ${\cal O} (N^2_f)$,
interpreted as that of a gas of (color singlet) mesons and
glueballs. The high-temperature (deconfinement) phase has a free
energy of order $N^2 f (N_f/N, \; T)$, which is interpreted as
that of a gas of gluons and of fundamental and anti-fundamental
matter states. This suggests the existence of a dual string
theory, and a transition to a black hole at  high temperature.
\end{quotation}

\newpage

 \renewcommand{\theequation}{1.\arabic{equation}}
 \setcounter{equation}{0}

\noindent{\large\bf 1. ~Introduction}

There has been steady interest in trying to find string duals to
large $N$ gauge theories at weak-coupling [1].  Although this is a
challenging problem, information keeps accumulating which keeps
the subject progressing [2--8].  Large $N$ gauge theories are
believed to undergo a deconfinement phase-transition at
sufficiently high temperature, which might possibly be related to
Hagedorn behavior in the dual string theory [9].  By considering
gauge theories on a compact space [10,11], one obtains an
additional parameter $R\L$ to vary which may be tuned to weak
coupling, where $R$ is the size of the compact space, and $\L$ is
the dynamical scale of the gauge theory. Because the gauge theory
is on a compact manifold, one must impose a Gauss law constraint,
which induces interactions among the gluons and the matter
multiplets. Aharony, {\it et al.} [11], have provided a general
framework, which we apply to the issues of our concern.

We are particularly interested in U(N) gauge theories with $N_f$
matter multiplets in the fundamental representation of the gauge
group, with $N_f/N$ finite in the large $N$ limit, as we have
previously discussed such theories [12] and proposed [7] for $d$=4
a possible infinite spin representation, bulk/boundary
correspondence, and $(\a^\prime )^{-1}$ expansion. In this paper
we show that these theories on $S^{d-1} \: \times$ time have two
phases, separated by a third-order phase transition at temperature
$T_c$. The free energy in the low-temperature phase behaves as
\be
 F/T \sim N^2_f \, f_1 (T) \hspace{1in} 0 \leq T \leq T_c
 \ee
and in the high-energy phase as
\be
 F/T \sim N^2 f_2 \left(\frac{N_f}{N} , \: T\right) \hspace{.75in} T \geq
 T_c \; .
 \ee
This is attributed to a gas of glueballs and (color singlet)
mesons in the low-energy phase, and a phase-transition to a
deconfinement transition, with the free-energy that of a gas of
gluons and fundamental and anti-fundamental matter states. It is
speculated that the low-temperature phase has a string dual, with
a high-temperature transition to a black hole.

In sec.\ 2 we specialize the work of Aharony, {\it et al.} [11],
to the models of our interest.  Section 3 explores the
phase-structure of these models.  The $d$=4 gauged vector models
[12] and their ${\cal N}$=1 supersymmetric cousins [7] are
presented in Sec.\ 4 as concrete examples of the theories we are
considering. Section 5 summarizes our principle findings, and
argues for the possible existence of string duals for the class of
theories studied throughout this paper.

\noindent{\large\bf 2. ~U(N) gauge theory on $\mathbf{S^{d-1}
\times} \sR$ at large N}

Aharony, {\it et al.} [11], showed that the partition function on
a compact space for free U(N) gauge theory, at large N, with free
matter multiplets, subject only to Gauss' law constraint is
 \renewcommand{\theequation}{2.\arabic{equation}}
 \setcounter{equation}{0}
\be
{\sZ} (x) = \int [dU] \exp \left\{ \sum_R \frac{1}{n} \left[ z^R_B
(x^n ) + (-1)^{n+1} z^R_F (x^n)\right] \cdot \chi_R (U^n)\right\}
 \ee
 In (2.1) $U$ is a $N \times N$ matrix, $\chi_R(U)$ is the
character for the representation $R$, $z^R_B(x)$ and $z^R_F (x)$
are the single-particle bosonic and fermionic partition functions
for each representation, and $x = e^{-\b} = e^{-1/T}$ at
temperature $T$.  [See refs.\ [10,11] for details of the
derivation, and for the explicit single-particle partition
functions.]  It is important to note that the Gauss' law
constraint induces interactions between the matter fields and the
gauge sector, even though the theory is at zero-coupling.

\noindent{\bf 2.1 ~U(N) gauge theory with $N_f/N$ finite}

Consider U(N) gauge theories with $N_f$ matter multiplets in the
fundamental representation of the gauge group, with $N_f/N$ finite
as $N\rightarrow \infty$.  Our motivation for discussing such
theories is to explore the possible bulk/boundary duality for such
field theories in the weak-coupling limit.  The putative bulk
string theory, considered in an $(\a^\prime )^{-1}$ expansion,
would correspond to the field theory at or near the UV
fixed-point.  It is speculated that the bulk theory has stringy
behavior, and thus studying the free-field theory on a compact
manifold may give additional insights for this issue.

With these considerations in mind, we restrict (2.1) to the
adjoint and fundamental representations.  Let $z_B(x)$ and
$z_F(x)$ be the single-particle partition function for the adjoint
representation, and ${\cal Z}_B(x)$ and ${\cal Z}_F(x)$ that for
the fundamental representation.  Then (2.1) becomes
\begin{eqnarray}
\sZ (x) & = & \int [dU] \exp \left\{ \sum^\infty_{n=1} \frac{1}{n}
\left[ {\cal Z}_B (x^n) + (-1)^{n+1} {\cal Z}_F (x^n)\right] \cdot
\Big[ tr (U^n) + tr (U^{+n}) \Big] \right. \nonumber\\
 & + & \left. \sum^\infty_{n=1} \frac{1}{n}
\left[ z_B (x^n) + (-1)^{n+1} z_F (x^n)\right]  tr (U^n) tr
(U^{+n}) \right\} \; .
\end{eqnarray}
It is to be emphasized that the same matrix $U$ appears for the
adjoint and fundamental matter, as the Gauss' law constraint has
induced interactions between these two sectors even at
zero-coupling.  One can write (2.2) in terms of the eigenvalues of
$U$; $ \{ \exp i\a_i \} \; (-\pi < \a_i \leq \pi )$ in the
standard way. That is,
 \be
  {\sZ} (x) = \int [d\a_i ] \exp - \left[ \sum_{i\neq j} V_A
  (\a_i-\a_j ) + \sum_i V_F (\a_i ) \right]
  \ee
  where [11]
$$
V_A(\th )  =  -\ell n |\sin (\th /2)| -  \sum^\infty_{n=1}
\frac{1}{n}\: \Big[ z_B (x^n) + (-1)^{n+1} z_F
 (x^n)\Big] \cos n\th
 \eqno{(2.4{\rm a})}
 $$
$$
= ~ \ell n \; 2 + \sum^\infty_{n=1} \frac{1}{n}\: \Big[1- z_B
(x^n) - (-1)^{n+1} z_F
 (x^n)\Big] \cos n\th~~~~~~
 \eqno{(2.4{\rm b})}
$$
and
 \renewcommand{\theequation}{2.\arabic{equation}}
 \setcounter{equation}{4}
\be
V_F(\th ) =  - \sum^\infty_{n=1} \frac{2}{n}\: \Big[ {\cal Z}_B
(x^n) + (-1)^{n+1} {\cal Z}_F  (x^n)\Big] \cos n\th \; .
\hspace{.9in}
 \ee
The second term in (2.4a), as well as (2.5) serve to bring the
eigenvalues closer. Thus, the presence of matter in the
fundamental representation increases the clustering of the
eigenvalues.

Introduce the eigenvalue distribution $\rho (\th )$ proportional
to the density of eigenvalues $e^{i\th}$ of $U$ at $\th$, with
$\rho$ everywhere non-negative subject to the choice of
normalization
\be
\int^\pi_{-\pi} d \th \rho (\th ) = 1  \; . \ee
 Define the moment
 \be
 \rho_n = \int^\pi_{-\pi} d \th \rho (\th )\cos n \th
 \ee
 where the eigenvalue distribution is assumed to be symmetric
 about $\th = 0$, as $V_A (\th ) = V_A (-\th )$ and
$V_F (\th ) = V_F (-\th )$.  Then, for $N_f$ multiplets in the
fundamental representation, the effective action for the
eigenvalues becomes
\be
S [\rho (\th )] = \frac{N^2}{\pi } \sum^\infty_{n=1} \left[
(\rho_n)^2 V_n^A (T) + 2 \left( \frac{N_f}{N} \right) \rho_n \:
V^F_n (T) \right]
 \ee
 where
 \be
 V^A_n (T) = \frac{\pi}{n} \Big[ 1-z_B (x^n) - (-1)^{n+1} z_F
 (x^n)\Big]
 \ee
 and
 \be
 V^F_n (T) = -\frac{\pi}{n} \Big[ {\cal Z}_B (x^n) + (-1)^{n+1} {\cal Z}_F
 (x^n)\Big]
 \ee
 for the adjoint and fundamental representations respectively.
 That is
 \begin{eqnarray}
 \sZ & = & \sum^\infty_{n=1} \exp \left[ \frac{N^2_f}{\pi} \;
 \frac{(V^F_n)^2}{(V^A_n)} \right] \nonumber \\[.2in]
 & \cdot & \int [d\rho_n] \sum^\infty_{n=1} \exp \left\{ -
 \frac{N^2}{\pi} \: (V^A_n) \left[ \rho_n + \left(
 \frac{N_f}{N}\right) \left(\frac{V^F_n}{V^A_n}\right)\right]^2 \right\}\;
 ,
 \end{eqnarray}
which can be written as
 \begin{eqnarray}
 \sZ & = & \sum^\infty_{n=1} \exp \left[ \frac{N^2_f}{\pi} \;
 \frac{(V^F_n)^2}{(V^A_n)} \right] \nonumber \\[.2in]
 & \cdot & \int [d\tilde{\rho}_n]\exp  - \frac{N^2}{\pi} \sum^\infty_{n=1}
 (\tilde{\rho}_n )^2 V^A_n (T)
 \end{eqnarray}
 where
 \be
 \tilde{\rho}_n = \rho_n +
 \left(\frac{N_f}{N}\right)\left(\frac{V_n^F}{V_n^A}\right)\; .
\ee
 The integral in (2.12) is analogous to that for purely adjoint
matter, but with $\tilde{\rho}_n$ instead of $\rho_n$.

A saddle-point solution of (2.11) or (2.12) implies that
\be
\rho_n = -
\left(\frac{N_f}{N}\right)\left(\frac{V_n^F}{V_n^A}\right)\; . \ee
or equivalently
\be
\tilde{\rho}_n = 0 \; .
 \ee
From (2.10) we observe that
\be
V^F_1(T) < 0 \ee
 always, and
$$
V^A_1(T) > 0 \; ,  \eqno{(2.17{\rm a})}
$$
 if
$$
[z_B(x) + z_F (x)] < 1 \; ,  \eqno{(2.17{\rm b})}
$$
\renewcommand{\theequation}{2.\arabic{equation}}
\setcounter{equation}{17}
so that $\rho_n \geq 0$ if (2.17) is
satisfied.  One might think that the solution $\tilde{\rho}_n = 0$
is valid as long as (2.17) is satisfied.  However, (2.7) implies
\be
| \rho_n | \leq 1 \; .
 \ee
If we formally define $T_H$ (the Hagedorn temperature) by
 \be
  [z_B (x_T) +
z_F (x_T)] = 1 \; ,
 \ee
 then a necessary condition imposed by (2.14)--(2.18) is
 \be
 1 \geq \rho_1 > 0 \; ,
 \ee
 which occurs for $0 \leq T < T_H$.  The Hagedorn temperature is
 never reached by the saddle-point solution, as (2.20) implies
 that
 \be
 \Big[ z_B (x) + z_F (x) \Big] + \left(\frac{N_f}{N}\right)
 \Big[{\cal Z}_B (x) + {\cal Z}_F(x) \Big] \leq 1
 \ee
 so that (2.19) is not compatible with (2.21).  We shall see in
 the next section that a third-order phase transition to a new
 phase occurs even before the upper-limit in (2.21) is attained.

 Also noteworthy is that the free-energy
 \begin{eqnarray}
 F & = & -T \; \ell n \, \sZ \nonumber \\
 & \sim & N^2_f \; ,
 \end{eqnarray}
 when the saddle-point solution is valid, as one observes from
 (2.12) with $\tilde{\rho}_n = 0$, contrary to the
 low-temperature phase for models with matter only in the adjoint
 representation, where $F \sim {\cal O}(1)$.

\noindent{\large\bf 3. ~ Phase structure}

\renewcommand{\theequation}{3.\arabic{equation}}
\setcounter{equation}{0}

Let us consider the phase-structure of this class of models in
more detail. The solution for the eigenvalue distribution $\rho
(\th )$ can be obtained from the equilibrium conditions for a
matrix model with action
 \be
 S = N \sum^\infty_{n=1} \frac{1}{n} \left[ a_n \rho_n +
 \frac{N_f}{N} \, b_n \right]\Big[tr (U^n) + tr (U^{+n}) \Big]
 \ee
 which generalizes (5.22) of ref.\ [11], where we have defined
 \be
 a_n = \Big[ z_B (x^n) + (-1)^{n+1}z_F(x^n)\Big]
 \ee
 and
 \be
 b_n = \Big[ {\cal Z}_B (x^n) + (-1)^{n+1}{\cal Z}_F(x^n)\Big]\; .
 \ee
This model, with action
 \be
 S = N \sum^\infty_{n=1} \frac{1}{2n} \: c_n \left[ tr (U^n)
 + tr (U^{+n}) \right]
 \ee
has been solved in ref.\ [13]), where here
 \be
c_n = 2 \left[ a_n\rho_n + \frac{N_f}{N} \: b_n \right]\; .
 \ee
[The notation in ref. [13] is $\b_n$ instead of $c_n$.]  The
method used by Aharony, {\it et al.} [11], to treat $\rho_n$ and
$\rho(\th )$ independently, and solve for $\rho_n$ and $\rho(\th
)$ self-consistently, using
\be
\rho_n = \int^\pi_{-\pi} d\th \rho (\th ) \cos n \th
 \ee
 works here as well.

There are two phases for the model under consideration, denoted by
$A_0$ and $A_1$ in the notation of ref. [13].  The phase $A_0$ is
characterized by a distribution of eigenvalues which covers the
circle $-\pi < \th \leq \pi$ without gaps, while in the phase
$A_1$ the circle has a single gap with zero density.

\noindent{\bf 3.1 ~ The $A_0$ phase}

The eigenvalue distribution in phase $A_0$ is not constant, but is
\be
 \rho_{A_{0}} = \frac{1}{2\pi} \left[ 1 + \sum^\infty_{n=1} c_n
 \cos n\th \right]
 \ee
 for a density chosen to be symmetric about $\th = 0$.
 Self-consistency requires
 \be
 \rho_n = \frac{1}{2} \: c_n = \left[ a_n\rho_n + \frac{N_f}{N} \,
 b_n \right] \; .
 \ee
 That is
 $$
 (\rho_n)_{A_0} = \left( \frac{N_f}{N} \right) \:
 \frac{b_n}{1-a_n} \hspace{.6in} \eqno{(3.9{\rm a})}
 $$
 $$
 = - \left( \frac{N_f}{N} \right) \:
 \frac{(V_n^F)}{(V^A_n)} \; , \eqno{(3.9{\rm b})}
 $$
where we have used (2.9) and (2.10).  The eigenvalue distribution
in phase $A_0$ is identical to that obtained from the saddle-point
solution, {\it i.e.}, (2.14), so that $\tilde{\rho}_n = 0$ in
phase $A_0$.  The free-energy for this phase is given by (2.11)
and (2.22).

\renewcommand{\theequation}{3.\arabic{equation}}
\setcounter{equation}{9}

\noindent{\bf 3.2 ~ The $A_1$ phase}

The $A_1$ phase can be described by a generalization of Aharony,
{\it et al.} [11], eq'ns. (5.23){\it ff}.  Consider a solution for
$\rho (\th )$ where $\rho (\th ) \neq 0$ for $- \th_0 < \th <
\th_0$, and $\rho (\th )$ otherwise, where $\th_0 = \pi$ at the
$(A_0, A_1)$ phase boundary.  The solution for $\rho(\th )$ in the
$A_1$ phase is
\be
\rho (\th ) = \frac{1}{\pi} \: \sqrt{s^2 - \sin^2
(\textstyle{\frac{\th}{2}})} \; \left[ \sum^\infty_{n=1} Q_n \cos
\Big((n -\textstyle{\frac{1}{2}})\th \Big)\right]
 \ee
 where $s = \sin (\th_0/2)$, and
 \be
 Q_n = \sum^\infty_{\ell = 0} c_{n+\ell}\: P_\ell (\cos \th_0)
 \ee
 where $P_\ell (x)$ is the Legendre polynomial.  From eq'n. (5.15)
 of ref. [13], we have (where $M=1$ for the $A_1$ phase)
 \be
 Q = Q_0 + 2 \; .
 \ee
That is (since $c_0 = 0$)
\be
\sum^\infty_{\ell = 1} \left[ a_\ell \rho_\ell + \frac{N_f}{N} \:
b_\ell \right] \left[P_{\ell -1} (1-2s^2) - P_\ell (1-2s^2)\right]
= 1 \; . \ee
 Define
 \be
 G_\ell = a_\ell [P_{\ell-1} (1-2s^2) - P_\ell (1-2s^2)]
 \ee
 and
 \be
 D_\ell = b_\ell [P_{\ell-1} (1-2s^2) - P_\ell (1-2s^2)] \; .
  \ee
Then (3.13) becomes
\be
\vec{G} \cdot \vec{\rho} = 1 - \frac{N_f}{N} \: \sum^\infty_{\ell
= 1} D_\ell \ee

Define
\be
B^{n-1/2} (s^2) = \frac{1}{\pi} \int^{\th_0}_{-\th_0} d\th \left\{
\sqrt{s^2 - \sin^2 (\textstyle{\frac{\th}{2}})}\; \left[ \cos
\Big( (n -\textstyle{\frac{1}{2}})\Big) \th \right] \right\}\; .
 \ee
Therefore
\begin{eqnarray}
\rho_n & = & \frac{1}{2} \sum^\infty_{k=1} Q_k \left[
B^{n+k-1/2}(s^2) + B^{|n-k+1/2|}(s^2)\right] \\ [.1in]
 & = &
 \sum^\infty_{\ell=0}\:\sum^\infty_{k=1} \left[a_\ell\rho_\ell +
 \frac{N_f}{N} \: b_\ell \right] \left[ B^{n+k-1/2}(s^2) +
  B^{|n-k+1/2|}(s^2)\right] \cdot P_{\ell -k} (1-2s)
\end{eqnarray}
where the polynomials $B^{n+1/2}$ are defined by
\be
\sum^\infty_{n=0} B^{n+1/2} (x) z^n = \frac{1}{2z} \left[
\sqrt{(1-z)^2 + 4zx} \; + (z-1)\right]\; .
 \ee
Let
 \be
 R_{n\ell} = a_\ell\sum^\ell_{k=1} \left[ B^{n+k -1/2}(s^2) +
 B^{|n-k+1/2|} (s^2)\right] P_{\ell -k}
 (1-2s^2)
 \ee
 and
 \be
 C_{n\ell} = b_\ell\sum^\ell_{k=1} \left[ B^{n+k -1/2}(s^2) +
 B^{|n-k+1/2|} (s^2)\right] P_{\ell -k}  (1-2s^2) \; .
 \ee
Thus (3.18) becomes
\be
\rho_n =\sum^\infty_{\ell =1} \left[ R_{n\ell} \rho_n +
\frac{N_f}{N}\; C_{n\ell} \right] \ee
 That is
\be
R\vec{\rho} = \vec{\rho} - \frac{N_f}{N} \sum^\infty_{\ell =1}
C_{n\ell}\; .
 \ee
 The eigenvalue moments and angle $\th_0$ in the $A_1$ phase are
 determined by (3.16) and (3.24), with solution
\be
\vec{\rho} =  - \frac{N_f}{N} (R-I)^{-1} \; \vec{C}
 \ee
 where $\vec{C} = \sum^\infty_{\ell = 1} C_{n\ell}$.

It is difficult to solve these equations in general, so we
consider the simplified model with $a_n = b_n = 0$ for $n \geq 2$,
which should give a good qualitative description of the model.  In
that case consider the pair of equations
\be
\rho_1 = R_{11}\rho_1 + \frac{N_f}{N} \; C_{11} \ee
 and
 \be
 G_1\rho_1 = 1 - \frac{N_f}{N} \; D_1 \; .
 \ee
Explicitly
\be
R_{11} = a_1 (2s^2 - s^4) \ee
\be
C_{11} = b_1 (2s^2 - s^4) \ee
 and
\be
D_1 = b_1 (2s^2 ) \ee
\be
G_1 = a_1 (2s^2 ) \; . \ee
 Combining (3.25)--(3.30) we find in the $A_1$ phase
 \begin{eqnarray}
 \rho_1 & = & \left( 1 - \frac{s^2}{2} \right) \nonumber \\
 & = & \left( 1 - \frac{1}{2} \: \sin^2 \; \frac{\th_0}{2}\right) \; ,
 \end{eqnarray}
 with
\be
 (\rho_1)_{A{_1}} = \textstyle{\frac{1}{2}}
\ee
 and
\be
 a_1 + \frac{2N_f}{N} \: b_1 = 1
\ee at the $(A_0,A_1)$ boundary.
 However, in the $A_0$ phase
\be
 (\rho_1)_{A{_0}} = \left( \frac{N_f}{N} \right) \;
 \frac{b_1}{1-a_1}\; .
 \ee
Since the single-particle partition functions are continuous,
(3.34) holds at the boundary of the $A_0$ phase, and thus
 \be
 (\rho_1)_{A{_0}} = \frac{1}{2} ~~~~~~~ {\rm at~the~boundary.}
 \ee
Hence $\rho_1$ is continuous across the $(A_0,A_1)$ phase
boundary, as it must.  Equations (3.33) or (3.36) imply that the
$(A_0,A_1)$ phase transition occurs before the bound (2.21) is
reached.  The temperature $T_c$ at which the phase transition
takes place satisfies (3.34), {\it i.e.},
\be
\left[ a_1 (x_c) + \frac{2N_f}{N} \: b_1 (x_c) \right] = 1 \; .
 \ee
Since (3.37) implies $a_1 (x_c) < 1$, so that $T_c < T_H$, as
expected.

One can solve for $\rho (\th )$, and $s = \sin^2 \frac{\th_0}{2}$
for the $A_1$ phase in the simplified model.  From (3.27) and
(3.30)--(3.32) we obtain
\begin{eqnarray}
s^2 & = & \left[ \sin^2 \, \left(
\frac{\th_0}{2}\right)\right]_{A{_1}} \nonumber \\
& = & \left( 1 + \frac{N_f}{N} \: \frac{b_1}{a_1} \right) -
\left[\left( 1 + \frac{N_f}{N} \: \frac{b_1}{a_1} \right)^2 -
\frac{1}{a_1} \right]^{1/2}
 \end{eqnarray}
 with $s^2 = 1$ at the $(A_0,A_1)$ boundary, and $0 \leq s^2 \leq
 1$ throughout the $A_1$ phase.  From (3.10), with
\be
Q_1  =  2 \left[ a_1\rho_1 + \frac{N_f}{N} \: b_1 \right]  =
\frac{1}{s^2}
 \ee
 and $Q_n = 0$ for $n\geq 2$, one obtains
\be
\rho (\th ) = \frac{1}{\pi s^2} \left[ s^2 - \sin^2 \left(
\frac{\th}{2}\right) \right]^{1/2} \cos \left(\frac{\th}{2}\right)
\ee
 and
\be
\rho_n = \int^{\th_0}_{-\th_0} d\th \, \rho(\th ) \cos n \th \; .
 \ee

For the simplified model, the free-energy in phase $A_0$, obtained
from (2.19) and (3.9), is
\be
\frac{1}{T} (F)_{A_0} = \frac{N_f^2 \:
b^2_1}{(1-a_1)}~~~~~~~~~~~~~
 \ee
 $$
\stackrel{\longrightarrow}{bdy}\; \frac{N^2}{4}\; (1-a_1)
 \eqno{(3.43{\rm a})}
$$
 $$
 ~~ = ~ \frac{N^2}{2} \: \left( \frac{N_f}{N} \right) b_1 \eqno{(3.43{\rm b})}
 $$
where we have used (3.34).  On the other hand, in the $A_1$ phase
 $$
 \frac{1}{T} \: (F)_{A_1} = N^2 \left\{ - \left[ \frac{1}{2s^2} +
 \frac{1}{2} \: \ell n \, s^2 -  \frac{1}{2} \right] -
 \left(\frac{N_f}{N} \right) b_1\rho_1 \right\}\hspace{1.24in} \eqno{(3.44{\rm a})}
 $$
 $$
 = N^2 \left\{ - \left[ \frac{1}{2} \, \ell n \, s^2 +
\frac{3}{4} - s^2 + \frac{s^4}{4} \right] + (1-a_1) \left( 1-s^2 +
\frac{s^4}{4} \right) \right\} \; , \eqno{(3.44{\rm b})}
$$
\renewcommand{\theequation}{3.\arabic{equation}}
\setcounter{equation}{44}
 where we used (3.32) and (3.39).  Since $s\rightarrow 1$ at the
 boundary,
 \begin{eqnarray}
\frac{1}{T}\; (F)_{A_1} & {_{\stackrel{\longrightarrow}{bdy}}} &\;
\frac{N^2}{4}\; (1-a_1)
\nonumber \\
&=&  \frac {N^2}{2} \left( \frac{N_f}{N} \right) b_1 \; .
 \end{eqnarray}
 Therefore, the free-energy is continuous across the boundary as
 required.  One may verify from (3.42) and (3.44) that the
 phase-transition, of the Gross--Witten type [14], is third-order.
 Alternately, the arguments [and Fig.\ 1] of ref.
 [13] arrive at the same conclusion.

Consider (3.44b) in conjunction with (3.38).  With $(\frac
{N_f}{N})$ fixed for a given model, the free-energy $F_{A{_1}}
\sim N^2 \, f_2 \left(\frac{N_f}{N} , \, T\right)$, while recall
that $F_{A{_0}} \sim N_f^2\,f_1 (T)$, in the $A_0$ phase, with a
smooth third-order transition at the $(A_0,A_1)$ boundary as
indicated by (3.43), due to the relation between $a_1$ and $b_1$
at the boundary.

\noindent{\bf 3.2 Is there a Hagedorn transition?}

Formally define the Hagedorn temperature $T_H$ by (2.19), or
equivalently by
\be
a_1 (x_H ) = 1 \; ,
 \ee
 noting from (3.37), that $T_H > T_c$ for our class of models.
 From (3.38), we observe that $0 < s^2 < 1$ at $x_H$.  Further
 both (3.38) and (3.44) are non-singular at $a_1=1$.  Recall that $x
 = e^{-1/T}$, so that $a_1$ increases for $T > T_H$, with $a_1
 \rightarrow \infty$ as $x \rightarrow 1$.  But from (3.38)
 $a_1\rightarrow \infty$ implies $s\rightarrow 0$, with $(F)_{A_{1}}<
 0$ in this limit.

From Fig.\ 1 of ref.\ [13], with $\b_2 = 0$, corresponding to our
simplified model, observe that there is no additional
phase-transition for $\b_1 > 1$.  [The $\b_n$ in that Figure is
equivalent to our $c_n$, {\it c.f.} (3.5).]  One might consider
$\b_2$ and $\b_1 \neq 0$, {\it i.e.}, $c_2$ and $c_1 \neq 0$, with
\be
c_2 = 2 \left[ a_2 (x^2) \rho_2 + \frac{N_f}{N} \: b_2 (x^2)
\right] \; .
 \ee
We show that $c_2 (x) << c_1 (x)$ for the explicit models we shall
consider.  Then (7.4) of ref.\ [13] applies, and no further
 phase-transitions are anticipated, as is also seen in Fig.\ 1 of
 ref.\ [13].

 The gaussian fluctuations to the saddle-point solution of
 (2.11)--(2.12) provide corrections which are ${\cal O}(1)$, {\it
 i.e}, $1/N^2$ corrections to the ${\cal O}(N^2_f)$ free-energy in
 phase $A_0$.  These gaussian fluctuations to the free energy are
 of the form
 \begin{eqnarray}
 \frac{1}{T} \: \d F_1 & \sim & \ell n \: V^A_1 (T)\nonumber \\
& \sim & \ell n (1-a_1 ) \; ,
\end{eqnarray}
 which becomes
 \be
 \frac{1}{T} \: \d F_1 \sim \ell n \: b_1
 \ee
 at the $(A_0 , A_1)$ boundary.  The restriction (3.37) means that
 these fluctuations do not diverge, contrary to models with
 adjoint matter only.

 Therefore, there does not appear to be Hagedorn phase-transition
 in these  models, by which we mean a divergent partition function, and
 accompanying phase-transition.

 \noindent{\large\bf 3.3 ~High-temperature behavior}

 From eq'n. (5.17) and (B.12) of Aharony, {\it et al.} [11], one has
 \be
 z_i (x) \rightarrow 2{\cal N}_i \: T^{d-1} + {\cal O}(T^{d-2})
 \ee
 at very high temperatures, where ${\cal N}_i$ is the number of
 physical polarizations of the fields of the theory, with
 space-time dimension $d$.  The potentials $V^F_n$ and $V^A_n$
 become strongly attractive in the large temperature limit, and
 $\th_0 \rightarrow 0$ in that limit, {\it i.e.}, $s^2 \rightarrow
 0$.  Thus, at very high temperatures $\rho_n \rightarrow 1$ to
 leading order, and $\rho (\th )$ approaches a delta-function.
 Therefore, for the class of models we are considering,
 \begin{eqnarray}
 \lefteqn{F(x\rightarrow 1)} \nonumber \\
 & = & -2 N^2 \, T^d \zeta (d) \left\{ \left[ {\cal N}^A_B +
 \left(\frac{2N_f}{N} \right) {\cal N}^f_B \right] \right.
 \nonumber \\
 & + & \left( 1-\frac{1}{2^{d-1}}\right) \left. \left[ {\cal
 N}^A_F + \left( \frac{2N_f}{N} \right) {\cal N}^f_F \right]
 \right\}
 \end{eqnarray}
 where ${\cal N}^A_B ({\cal N}^f_B )$ and ${\cal N}^A_F ({\cal
 N}^f_F)$ are the bosonic and fermionic degrees of freedom for the
 adjoint (fundamental) fields respectively, so that free-energy
 behaves as $N^2$ for the adjoint and $NN_f$ for the fundamental
 representations.

\noindent{\large\bf 4.  ~Specific Models}

\renewcommand{\theequation}{4.\arabic{equation}}
\setcounter{equation}{0}

One of our primary interests is to examine the thermodynamic
structure of large $N$ gauged vector models in four-dimensions
[12], as these theories are candidates [7] for an $(\a^\prime
)^{-1}$ expansion in AdS$_5$, and a conjectured AdS$_5$/CFT
correspondence.  These theories have both UV and IR fixed points,
with the considerations of this paper relevant to the UV
fixed-point, while the IR fixed-points are also within the
perturbative domain for appropriately chosen $(N_f/N)$.

The single-particle partition functions for $d$=4 are
\be
z_S = \frac{x^2 + x}{(1-x)^3}
 \ee
\be
z_V = \frac{6x^2 + 2x^3}{(1-x)^3}
 \ee
 and
 \be
z_F = \frac{8x^{3/2}}{(1-x)^3}
 \ee
 for scalar, vector, and Dirac fermions respectively.  Identical
 single-particle partition functions apply to ${\cal Z}_S (x)$, ${\cal Z}_V
 (x)$, and ${\cal Z}_F (x)$ as well.

\noindent{\bf 4.1  ~Gauged vector model}

The gauged U(N) vector model [12] has a scalar and $N_f$ fermions
in the fundamental representation, taken in the large $N$ limit,
with $N_f/N$ finite.  [That is, the gauged vector model is coupled
to the Banks--Zaks model [15].]  There is a window [12],
\be
3.6 \simeq \left(\frac{3\sqrt{3}}{2} + 1 \right) \leq N_f/N \leq
\frac{11}{2}
 \ee
 for which there is an IR fixed-point, which is
 more restrictive than that of the Banks--Zaks model.

 For both the gauged vector-model, and the Banks--Zaks model
\be
 a_n (x) = z_V(x^n)
 \ee
 and
\be
 b_n (x) = (-1)^{n+1} {\cal Z}_F(x^n)
 \ee
in the large $N$ limit.  The $(A_0,A_1)$ phase-transition occurs
when
\be
 a_1 (x_c)  + \frac{2N_f}{N} \: b_1 (x_c) = 1
 \ee
 according to (3.34), or
 \be
 z_V (x_c)  + \frac{2N_f}{N} \: {\cal Z}_F (x_c) = 1 \; ,
\ee
 Choosing $(N_f/N)=5$, as in the figure of ref.\ [7], gives
\be
 x_c \simeq 0.048 \; .
 \ee
 Equation (4.9) justifies the use of the simplified model of
 (3.2).

\noindent{\bf 4.2  ~Supersymmetric gauged vector model}

Consider ${\cal N}$=1 supersymmetric QCD with gauge group SU(N),
$N_f$ chiral multiplets $Q^i$ in the fundamental representation,
$\tilde{Q}_{\tilde{i}}$ in the anti-fundamental representation
$(i, \tilde{i} = 1$ to $N_f$), and a massless chiral superfield
$\s$, which is a color and flavor singlet [7].  The chiral
superfields interact by means of the superpotential
\be
W = \sqrt{\frac{\l}{N}} \: \s \sum^{N_f}_{i=1}
Q^i\tilde{Q}_{\tilde{i}} \; .
 \ee
The model with $\l = 0$ was studied extensively by Seiberg [16].
For our discussion, we restrict consideration to the non-Abelian
Coulomb phase with $3N/2 < N_f < 3N$, which is the conformal
window.

For this model the $(A_0,A_1)$ phase-transition occurs when (4.7)
is satisfied.  Here
\be
a_1 (x) = z_V(x) + z_F(x)
 \ee
and
\be
b_1 (x) = 2{\cal Z}_S(x) + {\cal Z}_F(x) \; ,
 \ee
as each chiral multiplet has a complex scalar and a Weyl fermion.
For example, choosing $(N_f/N) = 2$, the $(A_0,A_1)$
phase-transition takes place when
\be
x_c \simeq 0.051 \; .
 \ee
 Once again the simplified model is justified.

\noindent{\large\bf 5. ~Discussion}

 \renewcommand{\theequation}{5.\arabic{equation}}
 \setcounter{equation}{0}

We have considered the thermodynamic phase structure for free U(N)
gauge theory together with $N_f$ free matter multiplets, at large
$N$ on a compact manifold (in particular $S^{d-1} \times$ time);
where the Gauss' law constraint induces interactions between the
gluons and the matter multiplets.  Two phases were found, with the
low-temperature phase exhibiting a free energy which behaves as
 \be
 F/T \sim N^2_f \, f_1 (T) \hspace{1in} 0 \leq T \leq T_c \; .
 \ee
There is a third-order phase transition to the second phase, for
which
\be
 F/T \sim N^2 f_2 \left(\frac{N_f}{N} \, , \, T\right) \hspace{.75in} T \geq
 T_c \; .
 \ee
for two different functions $f_1$ and $f_2$ of the indicated
variables, subject to
\be
N^2_f \, f_1 (T_c) = N^2 f_2 \left(\frac{N_f}{N} , \, T_c \right)
\; .
 \ee
 At very high temperatures, the limiting behavior of the free
 energy is given by (3.51), as appropriate for a gas of deconfined states.

 The low-energy phase of this class of theories is that of a gas
 of glueballs and (color singlet) mesons $M^b_a \:(a,b, = 1$ to
 $N_f$), since the glueball contribution gives an ${\cal O}(1)$
 contribution to the free energy, while the mesons have
 ${\cal O}(N^2_f)$  degrees of freedom.  The phase-transition at
 $T_c$ is a (third-order) deconfining transition, where for
 $T>T_c$ one has a gas of gluons, and fundamental and
 antifundamental matter states, since gluons have $N^2$ degrees of
 freedom, and $N_f$ fundamental matter states contribute ${\cal
 O}(NN_f)$ to the free-energy.  Our computation of the free
 energies in phases $A_0$ and $A_1$ supports this picture.

 The challenge is to find string duals for weakly coupled field theories.
It has been argued that large $N$ deconfined phases (our $A_1$
phase) should be associated with black holes [9], even for weakly
coupled gauge theories [11].  That as, for weakly coupled gauge
theories in the high temperature phase, one may search for a bulk
dual, with radius of curvature of the order of the string scale
$\ell_s$ where
 \be
 \left(\frac{\ell_p}{\ell_s} \right) << 1  \; .
 \ee
 Standard arguments [9,11] show that the entropy behaves as $N^2$ in the
 high temperature phase.

 Is there evidence of stringy behavior in the low-temperature
 $(A_0)$ phase?  In eq'n. (3.48) we argue that the ${\cal O}(1)$
 Gaussian fluctuations give a $1/N^2$ correction to free-energy
 which behaves as
 \begin{eqnarray}
 \frac{a}{T} \: \d F_1 & \sim & \ell n \: V^A_1 (T) \nonumber \\
 & \sim & \ell n \: (1-a_1) \; .
 \end{eqnarray}
We interpret this to mean that the glueballs contribute to the
density of states which goes as
\be
\rho_{\rm adj} (E) \sim \frac{1}{E} \: e^{E/T{_H}}\hspace{1in} 0
\leq T \leq T_c < T_H
 \ee
 which is ``stringy".  However, the ``limiting" temperature is
 never reached, as $T_c < T_H$, since there is a deconfinement
 phase-transition at $T_c$.  We associate the glueball states with
 closed strings.  On the other hand, one should associate the gas
 of meson states $M^b_a$ in the low-temperature phase and the
gas of fundamentals and anti-fundamentals in the high temperature
phase, with open strings.

Notice that the specific models discussed in Sec.\ 4 all have IR
fixed-points, which prevent extrapolation to arbitrary large
't~Hooft couplings, and thus exclude the limit of $R_{AdS} >>
\ell_s$, which is the case of the better understood Maldacena
limit [17]. For $d$=4 one might be tempted to consider a IIB model
with $N$ D3 branes and $N_f$ D7 branes, with $N_f/N$ of ${\cal
O}$(1) in the large $N$ limit, but this is not possible, as the
number of allowed D7 branes is limited. In short, a specific
string or brane picture eludes us.  [See however ref.\ [7] for
conjectured infinite spin-representations for examples in $d$=4.]

In conclusion, there is significant evidence that conformal
theories at weak coupling have string duals, but there is a great
deal that must be done to make this more concrete.

\noindent{\large\bf Note added:}

Earlier work related to this paper is [18], where one considers a
SU(N) colored, quark-gluon gas partition function for $d$=3 in the
large $N$ limit, with $N_f$ fundamentals, taking into account the
colored-single constraint.  It was shown that the first-order
phase transition of the pure gluon gas changes to a third-order
phase transition when $N_f/N$ is finite, but making use of
Boltzmann statistics only.  The general set-up is presented in
[19].  Extensions to a conserved U(1) baryon number current in
[20] may also be of interest. We thank Professor Skagerstam for
bringing these papers to our attention.

We also acknowledge the helpful suggestions of the referee.

\noindent{\large\bf Acknowledgements}

This research was supported in part by the DOE under grant
DE--FG02--92ER40706.  I wish to thank Shiraz Minwalla, Anton
Ryzhov,  and especially Albion Lawrence for very useful
conversations.  I am  grateful to the string theory group and
Physics Department of  Harvard University for their hospitality
extended over a long period of time.

\noindent{\large\bf References}
 \begin{itemize}
 \item[1.]
 E.S.\ Fradkin and M.A.\ Vasiliev, Annals of
Physics (NY) {\bf 177} (1987) 63; M.A.\ Vasiliev,  hep-th/9910096;
hep-th/0104246; hep-th/0106149; Phys.\ Lett.\ {\bf B567} (2003)
139, hep-th/0304049; E.\ Sezgin and P.\ Sundell, JHEP, {\bf 0109}
(2001) 036, hep-th/0105001;  Nucl.\ Phys.\ {\bf B644} (2002) 303,
Eratum {\bf B660} (2003) 403, hep-th/0205131; C.\ Fronsdal, Phys.\
Rev.\ {\bf D20} (1979) 848; Phys.\ Rev.\ {\bf D26} (1982) 1988;
M.\ Flato and C.\ Fronsdal, Lett.\ Math.\ Phys.\ {\bf 8} (1984)
159; S.\ Ferrara and C.\ Fronsdal, Class.\ Quant.\ Grav.\ {\bf 15}
(1998) 2153, hep-th/9712239; M.\ G\"{u}naydin and D.\ Minic,
Nucl.\ Phys.\ {\bf B523} (1998) 145, hep-th/9802047; M.\
G\"{u}naydin, Nucl.\ Phys.\ {\bf B528} (1998) 432, hep-th/9803138;
M.\ G\"{u}naydin, D.\ Minic, and M.\ Zagermann, Nucl.\ Phys.\ {\bf
B534} (1998) 96, Eratum {\bf B538} (1999) 531, hep-th/9806042; G.\
Mack, Comm.\ Math.\ Phys.\ {\bf 55} (1977) 1;  M.\ Flato and C.\
Fronsdal, Lett.\ Math.\ Phys.\ {\bf 8} (1984) 159; V.K.\ Dobrev
and V.B.\ Petkova, Phys.\ Lett.\ {\bf 162B} (1985) 127; S.\
Ferrara and A.\ Zaffaroni, Phys.\ Lett.\ {\bf B431} (1998) 49,
hep-th/9803060; hep-th/9807090; D.Z.\ Freedman, S.S.\ Gubser, K.\
Pilch, and N.P. Warner, Adv.\ Theor.\ Math.\ Phys.\ {\bf 3} (1999)
363, hep-th/9904017; A.\ Ceresole, G.\ Dall'Agata, R.\ D'Auria,
and S.\ Ferrara,   Phys.\ Rev.\ {\bf D61} (2000) 066001,
hep-th/9905226.
 \item[2.]
W.\ Muck and K.S.\ Viswanathan,  Phys.\ Rev.\ {\bf D60} (1999)
046003, hep-th/9903194; L.\ Girardello, M.\ Porrati, and A.\
Zaffaroni,   Phys.\ Lett.\ {\bf B561} (2003) 289, hep-th/0212181;
N.V.\ Suryanarayana,  JHEP {\bf 0306} (2003) 036, hep-th/0304208;
R.G.\ Leigh and A.C.\ Petkou, JHEP {\bf 0306} (2003) 011,
hep-th/0304217; hep-th/0309177; E.\ Sezgin and P.\ Sundell,
hep-th/0305040.
 \item[3.]
 A.C.\ Petkou, JHEP {\bf 0303} (2003) 049, hep-th/0302063; T.\
Leonhardt, A.\ Megiane, and W.\ Ruhl,  Phys.\ Lett.\ {\bf B555}
(2003) 271,  hep-th/0211092; T.\ Leonhardt and W.\ Ruhl,
hep-th/0308111.
\item[4.]
P.\ Haggi-Mani and B.\ Sundborg,  JHEP {\bf 0004} (2000) 031,
hep-th/0002189; B.\ Sundborg, Nucl.\ Phys.\ Proc.\ Suppl.\ {\bf
16} (2001) 113, hep-th/0103247; A.\ Mikhailov,  \linebreak
hep-th/0201019; E.\ Witten, talk at the John Schwartz 60$^{\rm
th}$ Birthday Symposium, \linebreak
http://theory.caltech.edu/jhs60/witten/1.html, Nov.\ 2001;  U.\
Lindstr\"{o}m and M.\ Zabzine, hep-th/0305098; G.\ Bonelli,
hep-th/0309222.
\item[5.]
G.\ Gopakumar,  hep-th/0308184.
\item[6.]
S.R.\ Das and A.\ Jevicki, Phys.\ Rev.\ {\bf D68} (2003) 044011,
hep-th/0304093.
\item[7.]
H.J.\ Schnitzer, hep-th/0310210.
\item[8.]
E.\ Witten, hep-th/0312171.
\item[9.]
E.\ Witten, Adv.\ Theor.\ Math.\ Phys.\ {\bf 2} (1998) 505,
hep-th/9803131.
\item[10.]
B.\ Sundborg, Nucl.\ Phys.\ {\bf B573} (2000) 349.
\item[11.]
O.\ Aharony, J.\ Marsano, S.\ Minwalla, K.\ Papadodinas, and M.\
Van Ramsdonk, hep-th/0310285.
\item[12.]
D.\ Olmsted and H.J.\ Schnitzer,  Nucl.\ Phys.\ {\bf B512} (1998)
237,  hep-th/9602069;  H.\ Rhedin and H.J.\ Schnitzer,  Nucl.\
Phys.\ {\bf B537} (1999) 516,  hep-th/9804008.
\item[13.]
I.\ Jurkiewicz and K.\ Zalewski, Nucl.\ Phys.\ {\bf B220} [FS8]
 (1983) 167.
\item[14.]
D.J.\ Gross and E.\ Witten, Phys.\ Rev.\ {\bf D21}  (1980) 446.
\item[15.]
T.\ Banks and A.\ Zaks, Nucl.\ Phys.\ {\bf B196} (1982) 189.
\item[16.]
N.\ Seiberg,  Nucl.\ Phys.\ {\bf B435} (1995) 129, hep-th/9411149.
\item[17.]
J.M.\ Maldacena, Adv.\ Theor.\ Math.\ Phys,\ {\bf 2} (1998) 231;
Int.\ J.\ Theor.\ Phys.\ {\bf 38} (1999) 1113, hep-th/9711200;
S.S.\ Gubser, I.R.\ Klebanov, and A.M.\ Polyakov, Phys.\ Lett.\
{\bf B428} (1998) 105, hep-th/9802109; E.\ Witten,  Adv.\ Theor.\
Math.\ Phys,\ {\bf 2} (1998) 253, hep-th/9802150.
\item[18.]
B.-S.\ Skagerstam, Z.\ Phys.\ {\bf C24} (1984) 97.
\item[19.]
B.-S.\ Skagerstam, Phys.\ A: Math.\ Gen.\ {\bf 18} (1985) 1.
\item[20.]
S.I.\ Azakov, P.\ Salmonsson, and B.-S.\ Skagerstam, Phys.\ Rev.\
{\bf D36} (1987) 2137.
 \end{itemize}

\end{document}